\documentclass[twocolumn]{revtex4-1}
\usepackage[latin9]{inputenc}
\usepackage{float}
\usepackage{amsmath}
\usepackage{amssymb}
\usepackage{graphicx}
\usepackage{esint}

\makeatletter
 
 \@ifundefined{textcolor}{}
 {%
   \definecolor{BLACK}{gray}{0}
   \definecolor{WHITE}{gray}{1}
   \definecolor{RED}{rgb}{1,0,0}
   \definecolor{GREEN}{rgb}{0,1,0}
   \definecolor{BLUE}{rgb}{0,0,1}
   \definecolor{CYAN}{cmyk}{1,0,0,0}
   \definecolor{MAGENTA}{cmyk}{0,1,0,0}
   \definecolor{YELLOW}{cmyk}{0,0,1,0}
 }

\usepackage{float}\usepackage{mathtools}

\makeatother

\begin{document}

\title{Coherent Diabatic Ion Transport and Separation in a Multi-Zone Trap
Array}

\author{R. Bowler}

\email{ryan.bowler@nist.gov}

\author{J. Gaebler}

\author{Y. Lin}

\author{T. R. Tan}

\author{D. Hanneke}

\altaffiliation{Current address: Department of Physics, Amherst College, Amherst, MA, 01002-5000 USA}

\author{J. D. Jost}

\author{J. P. Home}

\altaffiliation{Current address: Institute for Quantum Electronics, ETH Zurich, 8093 Zurich, CH}

\author{D. Leibfried}

\author{D. J. Wineland}

\begin{abstract}
We investigate the motional dynamics of single and multiple ions during
transport between and separation into spatially distinct locations
in a multi-zone linear Paul trap. A single $\ensuremath{^{9}{\rm {Be}^{+}}}$
ion in a $\sim$ 2~MHz harmonic well located in one zone was laser-cooled
to near its ground state of motion and transported $370\,\mu$m by
moving the well to another zone. This was accomplished in $8\,\mu$s,
corresponding to 16 periods of oscillation. Starting from a state
with $\bar{n}\thickapprox0.1$ quanta, during transport the ion was
excited to a displaced coherent state with $\bar{n}\thickapprox1.6$~quanta
but on completion was returned close to its motional ground state
with $\bar{n}\thickapprox0.2$. Similar results were achieved for
the transport of two ions. We also separated chains of up to 9 ions
from one potential well to two distinct potential wells. With two
ions this was accomplished in $55\,\mu$s, with final excitations
of $\bar{n}\thickapprox2$ quanta for each ion. Fast coherent transport
and separation can significantly reduce the time overhead in certain
architectures for scalable quantum information processing with trapped
ions.
\end{abstract}
\maketitle
For quantum information processing (QIP) based on trapped ions \citep{95Cirac},
scaling might be achieved with an array of interconnected trap zones
where information transport is accomplished by confining the ions
in potential wells that are moved between zones \citep{98Wineland,02Kielpinski}.
The basic features of this scheme have been demonstrated in the situation
where ion qubits were transported adiabatically, on a time scale much
greater than the period corresponding to the ions' oscillation frequencies
in their local wells. This suppresses undesired motional excitation
that can impede the ability to perform high-fidelity multi-qubit quantum
logic gates \citep{02Rowe,09Jost,09Home,10Hanneke,11Blakestad}. In
\citep{04Barrett}, one and two ions were separated from a group of
three in 2 ms with mode excitation less than one quantum. In \citep{09Jost,09Home,10Hanneke}
a mixed-ion linear chain or ``crystal'' of ions was separated into
two $^{9}$Be$^{+}$-$^{24}$Mg$^{+}$ pairs accompanied by motional
excitation that could be removed with sympathetic laser cooling. In
these latter experiments, the time required for ion separation, transport,
and sympathetic laser cooling was 100 times larger than for logic
gates (approximately 5 and 20~$\mu$s for single and two-qubit gates,
respectively), emphasizing the potential for improvement.

With the goal of reducing the time required for ion separation, transport,
and sympathetic laser cooling, we have investigated diabatic transport
and separation. In \citep{08Huber}, diabatic ion transport was accomplished
on a time scale corresponding to four oscillation periods, but this resulted
in large motional excitation. Diabatic transport with small final
excitation of the center-of-mass (COM) motion has been observed for
cold neutral-atom ensembles on 1~s timescales (on the order of a
single oscillation period) \citep{08Couvert}. Here we demonstrate
diabatic ion transport in a duration (8 $\mu$s) comparable to that
for gate operations, with significantly suppressed final excitation.
We also demonstrate the separation of two $^{9}$Be$^{+}$ ions from
a common harmonic well into spatially distinct wells in $55\,\mu$s
and with a final excitation of approximately two quanta.

\begin{figure}[H]
\begin{centering}
\includegraphics[width=9 cm]{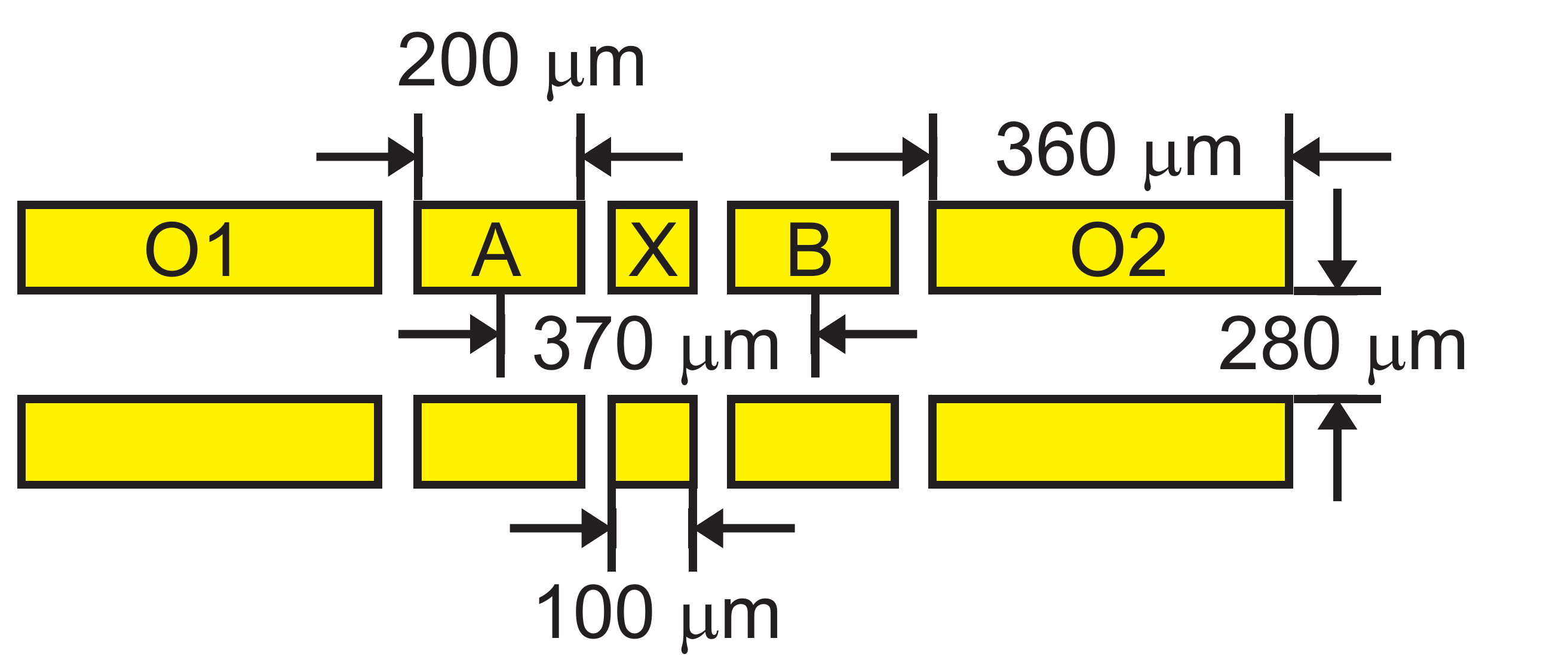}
\par\end{centering}

\caption{Schematic of the linear Paul (quadrupole) ion trap electrode structure
showing the two diagonally opposite segmented DC electrodes (not to
scale; trap details in \citep{09Jost,09Home,10Hanneke,10Jost}). Ions
are loaded in a region to the left of electrode~O1 and initially
transported to zone~A. Trap radio-frequency (RF) electrodes (not
shown) are referenced to a common ground potential.}

\label{fig:trappicture}
\end{figure}

Transport and separation of ions can be implemented by applying time-varying
potentials (waveforms) to segmented trap electrodes. In the experiments
described here, ions are confined in a multi-zone linear Paul trap
(Fig. \ref{fig:trappicture}) in which an ion's local potential well
can be made harmonic to a good approximation. To determine the total
potential at any given time in a waveform, we use simulations of potentials
from individual electrodes that are linearly superposed. Along the
trap axis (horizontal in Fig. \ref{fig:trappicture}), the confinement
is characterized by (angular) frequency $\omega$. Confinement in
the transverse direction is significanly stronger and does not play
a significant role here. The custom-built waveform electronics have
an update rate of 50~MHz, far above axial mode frequencies, enabling
considerable changes in potential within single periods of oscillation.
Zones~A, X, and B are regions near the respective electrodes through
which ions travel along the trap axis, while the outer electrodes
O1 and O2 are used only for potential optimization.

The qubit is formed from two 2s$^{2}$S$_{1/2}$ electronic ground-state
hyperfine levels of $\ensuremath{^{9}{\rm {Be}^{+}}}$ $|F=2,m_{F}=1\rangle\equiv|\downarrow\rangle$
and $|F=1,m_{F}=0\rangle\equiv|\uparrow\rangle$, separated in frequency
by $\omega_{0}/(2\pi)\simeq$ 1.2~GHz, where $F$ and $m_{F}$ are
the ion's total angular momemtum and its projection along the quantizing
magnetic field ($\simeq11.9$ mT). Ion motional quantum states can
be described in the Fock state basis $|n\rangle$ with energy $\hbar\omega(n+\frac{1}{2})$.
To characterize the motional states experimentally, we use two laser
beams detuned from each other by $\omega_{0}+\omega$ or $\omega_{0}-\omega$
to drive motion adding sideband (MAS) spin-flip transitions $|\downarrow,n\rangle\rightarrow$
$|\uparrow,n+1\rangle$ or motion subtracting sideband (MSS) transitions
$|\downarrow,n\rangle\rightarrow$ $|\uparrow,n-1\rangle$, respectively,
via stimulated-Raman transitions. Acousto-optic deflectors enable
the laser beams to be directed to either zone~A or B. Since the corresponding
Rabi rates $\Omega_{n,n\pm1}$ are $n$-dependent, the populations
${\rm {P}_{n}}$ of Fock states can be determined from the probability
${\rm {P}_{\downarrow}(t)}$ of the state $|\downarrow\rangle$ as
a function of drive time $t$, derived from state-dependent resonance
fluorescence \citep{96Meekhof}. As an approximation, we fit to the
function ${\rm {P}_{\downarrow}(t)=\frac{1}{2}(1+e^{-\gamma t}\sum_{n=0}{\rm {P}_{n}\cos(2\Omega_{n,n\pm1}t)})}$$ $
to include a phenomenological decay rate $\gamma$$ $.

Theoretical investigations of optimal transport of atomic systems
involve analyzing the acceleration and deceleration of an ion in a
way to minimize excitation of motion for a given transport duration.
\citep{06Reichle,06Schulz,08Huber,11Torrontegui,11Lau}. In our experiments,
the transport waveforms are designed to move the ions in a constant-curvature
harmonic potential characterized by $\omega$; if the ion
starts in the ground state, its quantum-mechanical motion will be
a displaced coherent state that follows a classical trajectory \citep{03Leibfried}.
A coherent state in such a well is characterized by a complex amplitude
$\alpha$ where $|\alpha|^{2}=\bar{n}$. For transport along z we
can write the potential as $\frac{1}{2}m\omega^{2}(z-z_{0}(t))^{2}$,
where $m$ is the mass of the ion and $z_{0}(t)$ is the minimum of
the well. In \citep{11Lau}, it has been shown that the coherent state
displacement $\alpha(t)$ resulting from transport in the frame of
the well center is

\begin{equation}
\alpha(t)=\sqrt{\frac{m\omega}{2\hbar}}\left(-e^{-i\omega t}\int_{0}^{t}\dot{z}_{0}(t')e^{i\omega t'}dt'\right).\label{Eq:AlpOft}
\end{equation}
 As long as the Fourier component of $\dot{z}_{0}(t')$ at the trap
frequency vanishes when integrated over the transport duration, the
ion will end up in its initial state at position $z_{0}(t)$. A simple
case where the well impulsively starts moving at constant velocity
$v$ and then stops at time $t_{T}$ ($\dot{{z}_{0}}=0$ for $t<0$
and $t>t_{T}$, and $\dot{{z}_{0}}=v$ for $t\in[0,t_{T}]$) leads
to

\begin{equation}
\alpha(t_{T})=\sqrt{\frac{m\omega}{2\hbar}}\left(i\frac{v}{\omega}\left[1-e^{-i\omega t_{T}}\right]\right).\label{Eq:AlpImp}
\end{equation}
 For $\omega t_{T}=2\pi N$, with $N$ an integer, an ion starting
in its ground state of motion is caught back in the ground state,
as in \citep{08Couvert}. For non-integer $N$, the ion will be left
in a coherent state different from the ground state. Due to the finite
response times of our electrode drive electronics, sudden starts and
stops are not practical. Nevertheless, Eq.(\ref{Eq:AlpOft}) gives
a criterion for transport functions $z_{0}(t)$ that leave the ion
in the ground state after the transport is completed. More generally,
one can always find a coherent displacement that returns the ion to
its ground state.

\begin{figure}
\includegraphics[scale=0.45]{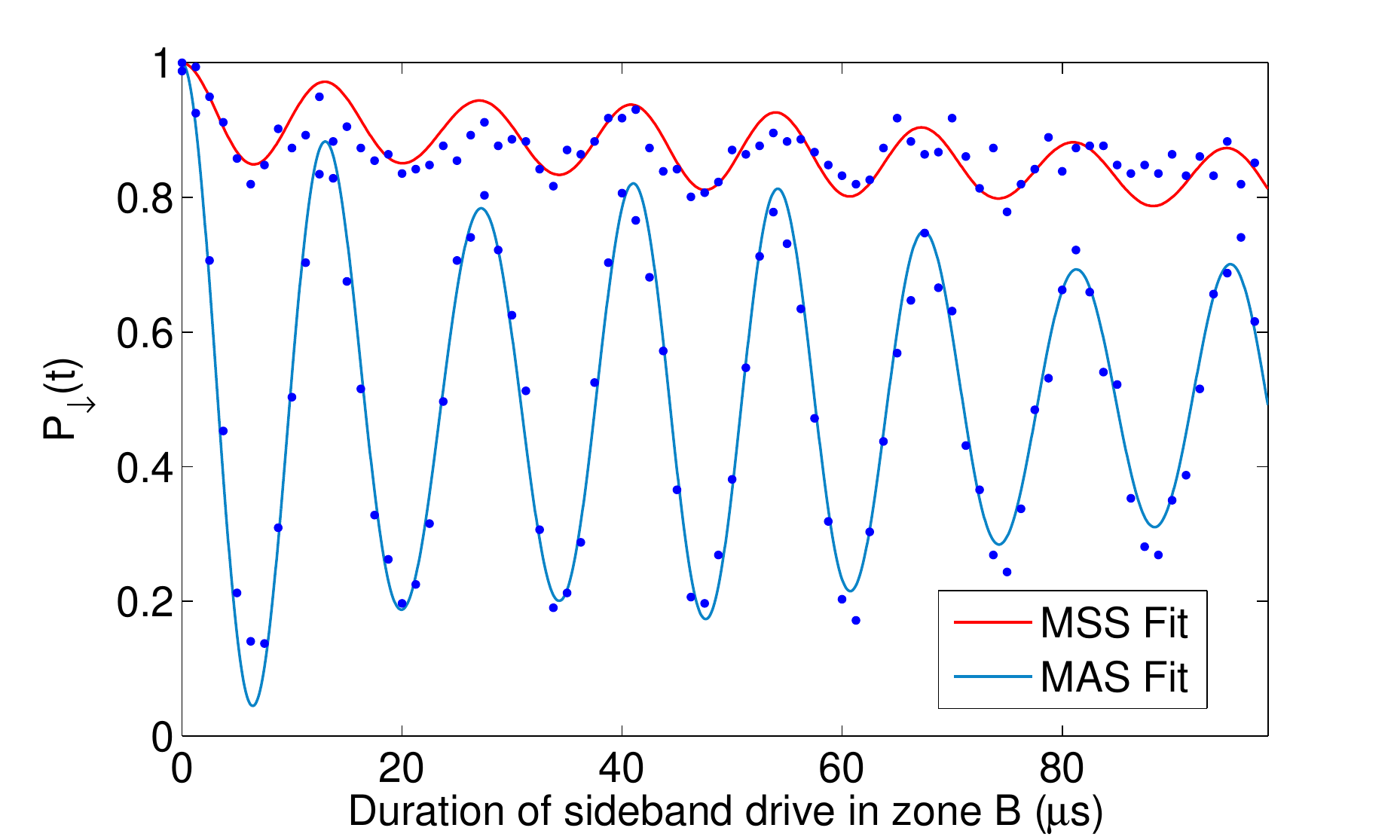}

\caption{Rabi flopping trace on the MAS and MSS after transport from zone~A
to B. Here we fit to thermal distributions that give $\bar{n}=0.19\pm0.02$
from the MAS fit and $\bar{n}=0.17\pm0.01$ from the MSS fit. The
axial frequency was $\omega/(2\pi)=1.972(1)$ MHz with a Lamb-Dicke
parameter $\eta=0.479$.}

\label{fig:OneWayColdFlop}
\end{figure}

As a demonstration of diabatic transport, a single ion in zone~A
($\omega/(2\pi)\simeq 2$ MHz; \{${\rm {V_{O1},V_{A},V_{X},V_{B},V_{O2}}}$\}
= \{1.289 0.327 2.173 0.310 1.311\}V) was first initialized; it was
laser-cooled to a thermal distribution with $\bar{n}\approx0.1$ and
optically pumped to $|\downarrow\rangle$. It was then transported
370 $\mu$m in $t_{T}\approx8\,\mu$s to zone~B (A $\rightarrow$
B). Following transport, qubit sideband transitions were driven in
zone B, then the ion was transported back to zone~A (B $\rightarrow$
A) to determine $P_{\downarrow}(t)$. By varying $\omega$, we minimized
the final excitation for $\omega/(2\pi)=1.972(1)$ MHz corresponding
approximatley to $N$ $=16$ (Fig. \ref{fig:OneWayColdFlop}). Excitation
was also minimized for values of $\omega/2\pi$ differing from this
value by integer multiples of $t_{T}^{-1}$.

\begin{figure}
\includegraphics[scale=0.45]{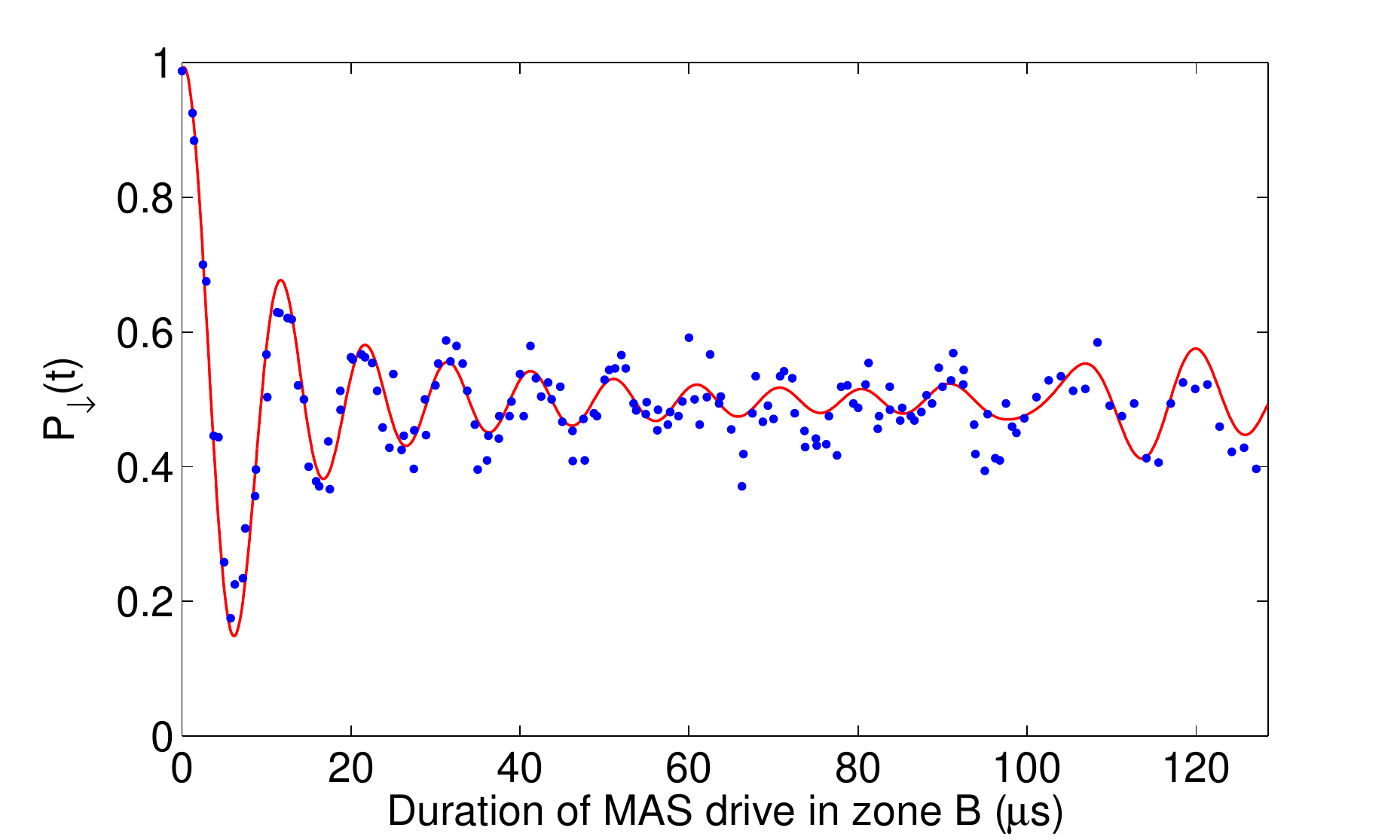}

\caption{Rabi flopping trace on the MAS following transport in which $\alpha(t_{T})\neq0$.
The figure shows a coherent state fit with $\bar{n}=6.4\pm0.2$, corresponding
to $|\alpha|=2.53\pm0.04$. The axial frequency was $\omega/(2\pi)=1.919(2)\,{\rm {MHz}}$
and the Lamb-Dicke parameter was $\eta=0.486$.}

\label{fig:OneWayDrivenFlop}
\end{figure}

For other values of $\omega/(2\pi)$, $P_{\downarrow}(t)$ measurements
showed that the final motional states were consistent with coherent
state probability distributions over Fock states; however, these measurements
do not verify the coherences between Fock states. To verify that the
transport excites a coherent state, we implemented A $\rightarrow$
B transport in a well of frequency $\omega/(2\pi)=1.919(2)$ MHz such
that $\alpha(t_{T})\neq0$ in zone~B (Fig.\ref{fig:OneWayDrivenFlop}).
We then applied a uniform electric field $E_{z}=E_{0}\cos(\omega t_{{\rm E}}+\phi_{E})$
to provide a phase-space displacement $\alpha_{E}$. This was followed
by a sideband drive of duration $t$, transport back to zone~A, and
measurement of $P_{\downarrow}(t)$. By adjusting $E_{0},t_{{\rm {E}}},$
and $\phi_{E}$ (relative to the time when transport started), we
could make $\alpha_{E}=-\alpha(t_{T})$. Under these conditions, we
measured a final state at $\bar{n}=0.19\pm0.02$. In addition, when
$\alpha(t_{T})\neq0$ in zone B, by waiting an appropriate delay
$T_{d}\pm N2\pi/\omega$ in B, and then transporting back to zone
A, the ion would be returned to its ground state in A (A $\rightarrow$
B $\rightarrow$ A ``cold'' transport).

Internal state qubit coherences are reliably maintained during transport
\citep{02Rowe,11Blakestad}. Under transport with minimal final excitation
we can expect that since any Fock state receives the same phase-space
displacement across the transport, initial motional states should
maintain the same relative coherences before and after transport.
To demonstrate this, we performed a Ramsey-type interference experiment.
After state initialization, a MAS $\pi/2$ pulse was applied to transform
the state to ${\textstyle {\frac{{1}}{\sqrt{{2}}}}}(|\downarrow,n=0\rangle+|\uparrow,n=1\rangle)$.
To establish a baseline reference, we first waited for a time equal
to the duration for A $\rightarrow$ B $\rightarrow$ A cold transport
followed by a second MAS $\pi/2$ pulse of variable phase. In this
case we measured a fringe contrast of 85(2) \%, limited primarily by
imperfect initial ground-state cooling and the relatively large Lamb-Dicke
parameter. When we performed the experiment with A $\rightarrow$
B $\rightarrow$ A cold transport between Ramsey pulses, the measured
86(2) \% contrast was consistent with no loss in coherence.

We also transported two $^{9}{\rm {Be}^{+}}$ ions (A $\rightarrow$
B) in the same well. If the moving potential well is maintained at
constant frequency, the ``stretch'' mode (ion motion out-of-phase)
should not be excited and the the COM mode should be excited as for
a single ion. We initialized both modes to $\bar{{n}}\approx0.1$
and after transport observed $\bar{n}_{COM}=0.35\pm0.1$, with negligible
excitation of the stretch mode.

When separating multiple ions confined in a single well into two separate
wells, it is impossible to preserve the motional mode frequencies.
To lowest order, the required external potential can be described
as the sum of a quadratic and a quartic term for $0\leq t\leq t_{s}$
during separation \citep{06Home}:
\begin{equation}
U(z)=a(t)z^{2}+b(t)z^{4},\label{eq:SepPotl}
\end{equation}
 with $a(0)>0$, $b(0)=0$ and $a(t_{s})<0$, $b(t_{s})>0$. Due to
Coulomb repulsion, the normal modes of multiple ions remain harmonic
throughout the separation for small excursions about each ion's local
minimum. The normal-mode frequencies go through their minima near
the point when $a$ vanishes and the quartic potential dominates the
confinement.

Our waveform for separation was divided into two main segments. Ions
were first transported to zone~X, whose center is defined as $z=0$
relative to Eq.(\ref{eq:SepPotl}). Then $a(t)$ was decreased and
$b(t)$ increased by lowering the potentials on electrodes~A and
B while simultaneously increasing the potentials on outer electrodes~O1
and O2 until the mode frequencies approximately reach their minimum
values. In the second segment, we separated the ions into distinct
wells in zones~A and B by increasing the potential on electrode~X
such that $a(t)$ changed sign, creating a ``wedge'' that splits
the ions apart.

\begin{figure}[H]
\begin{centering}
\includegraphics[scale=0.45]{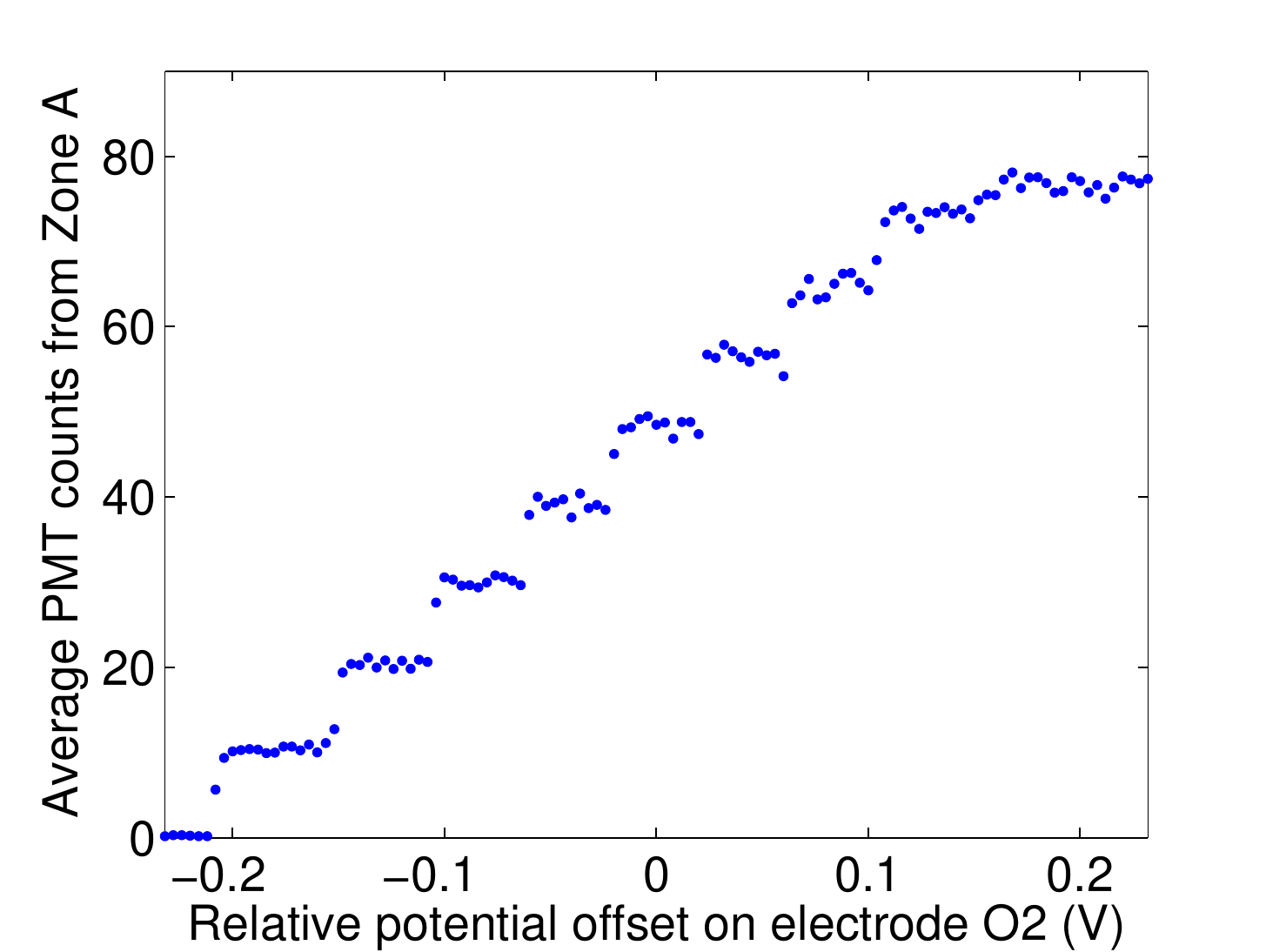}
\par\end{centering}

\caption{Ion fluorescence in zone~A as a function of the offset potential
on electrode~O2 during separation relative to a central value. Each
step increase of the fluorescence corresponds to the presence of an
additional ion in zone~A after separation. Any number of the nine
ions could be reliably separated into zone~A by setting the offset
potential to the center of the respective step. Fluorescence of a
single ion led to approximately 10 average photomultiplier tube (PMT)
counts in the detection period, but as more ions were added in zone~A,
the crystal was no longer uniformly illuminated by the detection beam
and the total fluorescence dropped below 10 counts per ion.}

\label{fig:9IonSeparation}
\end{figure}

As one demonstration of ion separation control, we ran a separation
waveform (duration $\approx340\,\mu$s) to partition a linear crystal
of nine $^{9}{\rm {Be}^{+}}$ ions into all possible combinations.
We first Doppler cooled the crystal in zone~A and optically pumped
to $|\downarrow_{1},\downarrow_{2},\cdots\downarrow_{9}\rangle$.
It was then transported it to zone~X for separation into two groups
in zones~A and B. A variable offset potential V$_{{\rm {O2}}}$ applied
to electrode~O2 imposed an electric field in zone~X that shifted
the center of the crystal relative to the separation wedge and determined
the number of ions in each group. Ions in zone~A were detected with
laser-induced fluorescence, approximately proportional to ion number.
Following detection, all ions were recombined and transported to zone~A
where the number was checked. As shown in Fig.\ref{fig:9IonSeparation},
increasing V$_{{\rm {O2}}}$ led to increased ion numbers in zone~A.
By adjusting V$_{{\rm {O2}}}$ to the center of a certain step, we
could reliably partition the ions into two groups of predetermined
number.

During separation, diabatic changes in both curvature and minimum
position of the axial potential can lead to coherent displacement
and squeezing of the motional states. The change in $\omega$ as well
as the accelerations of the potential well minima are both determined
by the rate of change in the separating wedge. The largest excitation
during separation coincides with the least adiabatic changes in mode
frequencies, near the time when the quadratic component of the potential
changes sign \citep{06Home}. These excitations can be suppressed
by approximating the adiabatic condition for ramping a harmonic well,
$\frac{1}{\omega^{2}}\frac{d\omega}{dt}\ll1$ \citep{10Chen}. We
numerically solved for this quantity in our waveform and imposed a
maximum value of $\frac{1}{\omega^{2}}\frac{d\omega}{dt}=0.025$ for
the COM mode during the ramp-down of the harmonic well and $\frac{1}{\omega^{2}}\frac{d\omega}{dt}=0.015$
around the sign change of $a$ in Eq.(\ref{eq:SepPotl}).

A two-ion crystal in zone~X with an initial well frequency of $\omega/(2\pi)=$
2.6~MHz (\{${\rm {V_{O1},V_{A},V_{X},V_{B},V_{O2}}}$\} = \{2.433,
-0.3763, -1.7089, -0.3831, 2.473\}V)could be separated and one ion
placed in zones~A and B (with motional frequencies 2.85~MHz and
2.77~MHz respectively, \{${\rm {V_{O1},...V_{O2}}}$\} = \{4.441,
-5.252, -0.649, -5.411, 5.952\}V) in a duration of $55\,\mu$s. After
separation, qubit MAS transitions of varying duration $t$ were driven
on either the ion in zone~A or B followed by recombination in zone
A, whereupon fluorescence state detection was performed to determine
$P_{\downarrow}(t)$ for each ion. The first separation segment took
$17\,\mu$s, during which it was necessary to ramp V$_{O2}$ from
2.473 to 5.952~V to compensate for trap geometric asymmetries and
to center the crystal over the separation wedge; the final value was
set by minimizing $\bar{n}$ for both ions. Near the point of ion
separation, simulation predicted that the minimum COM mode frequency
was 700~kHz and minimum stretch mode frequency was 880~kHz. During
the second separation segment we obtained best results by also applying
a variable offset potential to electrode~X to tune the quartic component
$b(t)$ in the potential. We also applied a relative difference between
electrodes~A and B to compenstate for trap imperfections and make
the potential well as symmetric as possible about electrode~X. This
enabled a fine-tune balancing of the final excitation between the
two ions. Experimentally, we found that equal excitation in both final
zones gave the lowest total excitation to both ions.

\begin{figure}
\includegraphics[scale=0.45]{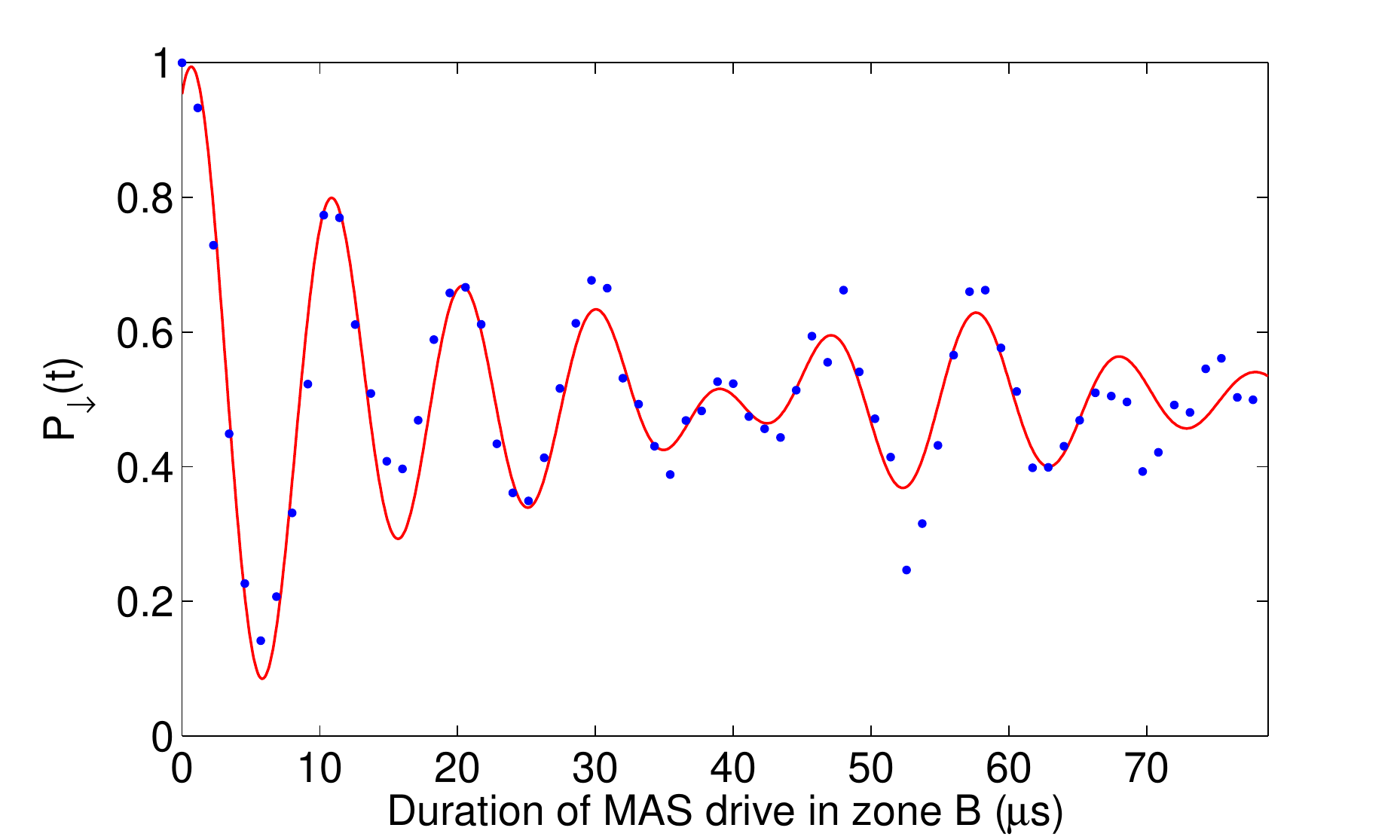}

\caption{Rabi flopping trace of the MAS of the ion in zone B after separation.
The fits are to a coherent state with $\bar{n}=1.9\pm0.1$, and $|\alpha|=1.38\pm0.04$
($\eta=0.404$). The trace for the ion in zone A looks very similar
giving $\bar{n}=2.1\pm0.1$, or $|\alpha|=1.45\pm0.03$ ($\eta=0.399$). }

\label{fig:separation_flopping}
\end{figure}

The two-ion crystal was initialized in zone X to a thermal state with
$\bar{n}\approx0.1$ on both the COM and stretch modes and optical
pumping to $|\downarrow_{1}\downarrow_{2}\rangle$. After separation,
sideband excitation gave Fock state populations consistent with coherent
states of $\bar{n}=2.1\pm0.1$ in zone~A and $\bar{n}=1.9\pm0.1$
in zone~B (Fig.\ref{fig:separation_flopping}). With RF de-excitation
of each ion, we could reduce $\bar{n}$ to $1.4\pm0.1$ and $1.6\pm0.1$
in zones~A and B respectively, consistent with a thermal distribution.
The reductions indicated partial coherence of the final states. The
residual excitations could not be explained by ambient heating; we
measured the heating rates of a single ion over the range of mode
frequencies experienced by the ions during separation and determined
that the integrated heating rates would lead to an additional $\Delta\bar{n}\approx0.2.$
The fact that we couldn't reduce $\bar{n}$ further could be explained
if the phase of the coherent states changed significantly from experiment
to experiment. This seems plausible, given the sensitivity of the ions'
local wells to slight changes in electrode potentials at the point
when the local potential curvatures are near their minima. As an example,
for a change of V$_{O2}$ of + 3~mV, the motion of the ion in zone~A
had $\bar{n}\simeq1.14\,\pm\,0.07$, while the ion in zone~B had $\bar{n}>15$,
which then could be coherently de-excited to a state consistent with
a thermal distribution having $\bar{n}=3.8\pm0.5$. The large excitation
of the ion in B can be qualitatively explained; since it is closer
to the maximum of the wedge during separation, it would be expected
to gain more kinetic energy as it accelerates towards the center of
its respective well.

In summary, we have demonstrated diabatic transport of one and two
ions and separation of two ions in a multi-zone linear trap on $10\,\mu$s
time scales, which approach those of logic gate operations. When transporting
and separating ions on a time scale that approached the ions' local
well oscillation periods, we observed coherent excitation of the motion
that could be avoided by shaping the waveform appropriately or nearly
eliminated by a coherent displacement that was applied after transport
or separation. In addition, multi-ion crystals were reliably partitioned
into two groups of predetermined numbers. These methods can reduce
the time overhead for separation, transport, and re-cooling in scalable
implementations of large quantum information algorithms.

The spatial extent of the separating ``wedge'' is governed by the
ion-to-electrode distances; therefore, smaller traps can accomplish
fast separation while maintaining stronger confinement, which would
lead to reduced excitation if background motional heating can be suppressed.
In principle it should be possible to transport like ions between
two zones in arbitrarily short durations \citep{02Wineland}; however,
when potential curvatures at the ion positions are time-dependent,
squeezing must also be taken into account for both transport and separation
(In the experiments reported here, squeezing was estimated to be negligible).
Moreover, experiments that use a different species for sympathetic
laser cooling \citep{09Jost,09Home,10Hanneke} will require special
consideration since all modes will be excited during both diabatic
transport and separation.

This work is supported by IARPA, NSA, ONR, DARPA \& the NIST Quantum
Information Program. We thank J. Heidecker for electronics support
and D. Slichter and A. Wilson for helpful comments on the manuscript.
Contribution of NIST; not subject to U.S. copyright.

\bibliographystyle{prsty}

\end{document}